\documentclass[12pt]{iopart}

\usepackage{graphicx}

\newcommand{\bx}{\mathbf{x}}

\newcommand{\bu}{\mathbf{u}}

\newcommand{\la}{\langle}
\newcommand{\ra}{\rangle}

\newcommand{\beq}{\begin{equation}}
\newcommand{\eeq}{\end{equation}}
\newcommand{\beqn}{\begin{eqnarray}}
\newcommand{\eeqn}{\end{eqnarray}}
\newcommand{\pp}{\partial}
\newcommand{\dd}{{\rm d}}
\newcommand{\ee}{{\rm e}}
\newcommand{\eq}{equation\ }
\newcommand{\cf}{c.f.\ }
\newcommand{\fig}{fig.\ }

\begin{document}

\title[Length distribution of stiff polymers ]{Length distribution of stiff, self-assembled polymers at thermal equilibrium}

\author{Chiu Fan Lee}

\address{Department of Bioengineering, Imperial College London
\\
South Kensington Campus, London SW7 2AZ, U.K.}
\ead{c.lee@imperial.ac.uk}
\begin{abstract}
We investigate the 
length distribution of self-assembled,  long and stiff polymers at thermal equilibrium.  Our analysis is based on calculating the partition functions of stiff polymers of variable lengths in the elastic regime. Our conclusion is that the length distribution of this self-assembled system follows closely the  exponential distribution, except at the short length limit. We then discuss the implications of our results on the experimentally observed length distributions in amyloid 
fibrils.
\end{abstract}
%\pacs{64.75.Yz}{Self-assembly}
%\pacs{82.35.Pq}{Biopolymers, biopolymerization}
%\pacs{87.14.em}{Fibrils (amyloids, collagen, etc.)}
\maketitle

\section{Introduction}
Biopolymers, such as microtubules, actin filaments and amyloid fibrils, differ from synthetic polymers in that the persistence lengths are usually greater than the
typical lengths of the biopolymers in the system. For instance, the persistence lengths, $l_p$, of microtubules, actin filaments, sickle cell hemoglobin fibers and amyloid fibrils have been estimated to be in the order of $10^2-10^3$ $\mu m$\footnote{Note that in the case of micotubules, the persistence length has been shown to demonstrate to be dependent on the total length of the biopolymers in \cite{pampaloni}} \cite{pampaloni}, 16 $\mu m$ \cite{legoff}, $10^3$ $\mu m$   \cite{wang} and $1-20$ $\mu m$ \cite{knowles2,adamcik,castro}, respectively. Another distinguishing feature of biopolymers from their synthetic counterparts is their relatively low polymerisation binding energies. For example, the binding energy for some amyloid fibrils have been estimated to be in the order of $-10$ kcal/mol \cite{onuallain,lee1}. This relatively low binding energy implies that thermal equilibrium may be reachable within experimentally relevant time scale.
 At thermal 
equilibrium, the polymerisation process will reach a steady state and the length distribution of the biopolymers constitutes one of the defining characteristics of the %%@
system. In particular, the knowledge of the length distribution provides an estimate on the binding energy, and thus also the stability, of the biopolymer concerned. In the dilute regime where pairwise interactions between polymers can be ignored, mean-field theory predicts that 
the length distribution at thermal equilibrium is exponentially distributed \cite{israelachvili,cates}. Specifically, if $\phi_L$ denotes the concentration of polymers of length $L$, then mean-field theory indicates that
\beq
\phi_L \propto \ee^{- L/\alpha}
\eeq
where  $\alpha = \la L \ra$ is the average length. In the mean-field approximation, each monomer within the polymer is considered identical. This is a coarse approximation as the contribution of each monomer to the partition function can potentially be depend on its position within the polymer. For instance, for long and flexible polymers, 
the above formula is modified due to the intra-chain volume exclusion interactions \cite{schoot,khokhlov}. Is the mean-field prediction also modified for stiff polymers in the regime $l_p > \la L \ra$? This is the question asked in this paper. Our analysis is based on mapping the calculations of the partition functions for stiff polymers to a quantum mechanical problem through the path integral formalism. Although this mapping has been employed extensively in the literature (e.g., see \cite{saito,schulman, stepanow,hamprecht, nakamura}), to the best of the authour's knowledge, this technique has never been employed to investigate the length distribution of self-assembled polymers.

Besides the intrinsic importance in understanding the thermodynamic properties, this work is also motivated by the recent experiments on amyloid fibrillisation in which a variety of
length distributions of amyloid fibrils formed from different proteins are observed \cite{Rogers_Macromol05,Rogers_EPJE05,vanRaaij_BiophysJ08,Xue_JBC09,Xue_ProtEngDesignSelect09}. 
 We will comment on  the implications of our results on these experimental %%@
observations in Section 3.

\section{Model}
To calculate the length distribution, we minimize the free energy of the system with respect to the length distribution within the saddle-point approximation. In the dilute limit where interactions between polymers are  negligible, we can write the total partition function as follows \cite{Mutaftschiev_B01}:
\beq
\label{Ptot}
Z_{\rm tot} =\prod_s' \frac{(Z_s)^{N_s}}{N_s!}
\eeq
where $N_s$ is the number of $s$-mers in the system. The prime in the product denotes the constraint
\beq
\label{constraint}
\sum_{s=1}^\infty s N_s =N
\eeq
where $N$ being the total number of monomers. In \eq (\ref{Ptot}), $Z_s$ denotes the partition that corresponds to a $s$-mer. Specifically, 
\beq
\label{Zs_def}
Z_s = \frac{1}{s! \ \Lambda^{3s}} \int_{\Gamma_s} \dd \bx_1 \cdots \dd \bx_s \ee^{-H(\{ x\}) / k_BT} \ ,
\eeq
where $H\{ x \}$ is the internal energy corresponding to the configuration $\{x\}$. The domain of integration, $\Gamma_s$, is constrained in such a way that
 the configuration domain represents a $s$-mer.  In \eq (\ref{Zs_def}), we have also integrated out the kinetic part of the partition function and so $\Lambda = h/\sqrt{2\pi %%@
m k_BT}$ corresponds to the thermal wavelength of the monomer \cite{Mutaftschiev_B01}. Since there are $(s!)$ ways of arranging the monomers within the $s$-mer, it cancels %%@
with the factor $(s!)$ in the denominator in \eq (\ref{Zs_def}) once the enumeration of the monomers within the polymer is fixed. To ease notation, we will also set $k_BT$ to one from now on.

Given the total partition in \eq (\ref{Ptot}), one can obtain the size distribution in thermal equilibrium by minimising the total free energy $F = -k_BT \ln Z_{\rm tot}$ with the constraint in \eq (\ref{constraint}) enforced by the Lagrange multiplier method. The resulting distribution is then of the form
\beq
\label{phis}
\phi_s \propto Z_s \ee^{-s \lambda}
\eeq
where $\phi_s$ is the concentration of $s$-mers in the system, and $\lambda$ is the Lagrange multiplier that enforces the conservation of the total monomer number \cite{cates, lee2}. 
Let $L_s$ be the length of the $s$-mer, we will from now on set the unit of length to $L_s/s$. As a result, $\phi_s$ is equivalent to the length distribution in the system.

Within the mean-field approximation, the integration in \eq (\ref{Zs_def}) is approximated as a product of $s$ identical integrals and so we have $Z_s = VKA^s$ for 
some constants $A$ and $K$, and $V$ corresponds to the total volume of the system. From \eq (\ref{phis}), the mean-field approximation therefore predicts that the length distribution is 
\beq
\label{mf}
\phi_s \propto \ee^{-s (\lambda -\ln A)}
 \ .
\eeq
Namely, the length distribution is exponential, a well known mean-field results \cite{cates,sciortino,lee2}.

%%%%%%%%%%%%%%%%%%%%%%%%%%%%%%%%%%%%%%%%%%%%%%%%%%%%%%%%
\begin{figure}
\begin{center}
\includegraphics[scale=.6]{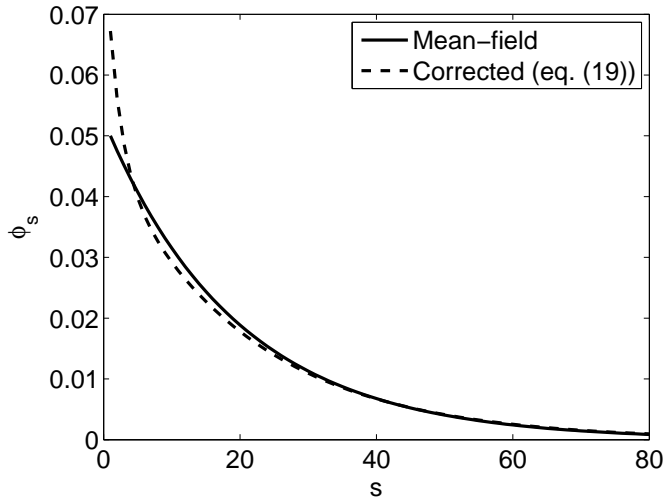}
\end{center}
\caption{The length distributions of self-assembled stiff polymers in thermal equilibrium according to the mean-field theory (\cf \eq (\ref{mf})) and the corrected version computed in this work (\cf \eq (\ref{new_res})). The average aggregation numbers, $\la s \ra$, are fixed to 20 in both cases. For the broken curve, $\kappa =  1$ and $\epsilon=10$. For parameters relevant to amyloid fibrillisation such that $\kappa, \epsilon, \la s \ra \gg 1$, the mean-field and the corrected results will be completely indistinguishable.
}
\label{plot}
\end{figure}
%%%%%%%%%%%%%%%%%%%%%%%%%%%%%%%%%%%%%%%%%%%%%%%%%%%%%%%

We will now go beyond the mean-field approximation. To do so,  one needs to account for the differential contributions to the $s$-mer partition function from different configurations of the $s$-mer. 
For long and stiff polymers, one can describe the energy of the polymer by its elastic constants for  bending ($\epsilon$) and for stretching ($\kappa$) \cite{harris,saito,soda}. This picture entails the assumptions that the elastic constants are uniform throughout the polymers and are length independent. Based on this continuum approach,  the partition function for a stiff polymer of length $s$ is  \cite{harris,saito,soda}
\beq
\label{eq_Zs}
Z_s = K
 \int_{\Gamma_s} D[\bx] \ee^{-H_s}
\eeq
where $D[\bx] \equiv \prod_k \bx_k$ and 
$K$ is some constant that does not depend on $s$ and will thus be ignored from now on. In \eq (\ref{eq_Zs}),
\beq
\label{stiffE}
H_s = (s-\gamma)\mu+\frac{1}{2} \int_0^{s} \dd t\left\{ \epsilon \left( \frac{\pp \bu}{\pp t} \right)^2
+\kappa [\xi(t)-1]^2 \right\} 
\ .
\eeq
The first term in the above Hamiltonian is responsible for the self-assembly process, where $\mu <0 $ accounts for the binding energy per monomer, and $\gamma \ll s$ corresponds to the missing binding energy at the two ends of the polymer \cite{cates,lee2}.
Furthermore, $t$ denotes the contour length of the polymer, whose configuration is given by $\bx(t)$. We also have $\bu(t) \equiv\pp  \bx /\pp t$ and $\xi(t) \equiv |\bu(t)|$. Note that the bending energy should strictly be dependent on the stretching as well, but the current form of the energy is valid for small elongation and contraction \cite{soda}. 

The integral in \eq (\ref{eq_Zs}) can be seen as a path integral in the quantum mechanics setting \cite{saito,schulman}. Specifically, if we fix the tangent vectors at the end points of the polymer in the configuration integral in \eq (\ref{eq_Zs}), then the path integral describes the evolution of the wavefunction, $\psi_s(t,\bu)$, that satisfies the following Schr\"{o}dinger-like equation:
\beqn
\frac{\pp \psi_s}{\pp t} &=& \frac{1}{2 \epsilon} \nabla^2_{\rm \xi} \psi_s - \frac{\kappa}{2} (\xi-1)^2 \psi_s
\\
\nonumber
&=& \frac{1}{2 \epsilon} \Bigg[ \frac{1}{\xi} \frac{\pp^2}{\pp \xi^2} (\xi \psi_s) + \frac{1}{\xi^2\sin \theta} \left(\sin \theta \frac{\pp \psi_s}{\pp \theta} \right) 
+
\frac{1}{\xi^2\sin^2 \theta} \frac{\pp^2 \psi_s}{\pp \phi^2}  \Bigg]
- \frac{\kappa}{2} (\xi-1)^2 \psi_s
\ ,
\eeqn
where $\bu$ is expressed in terms of the polar coordinates $( \xi, \theta, \phi)$. The Green's function for the above differential equation is known  \cite{saito}:
\beqn
\label{Gf}
 G(t,\xi, \theta, \phi| t',\xi', \theta', \phi') &=& \sum_{p,n,m} \exp \left[ -\frac{\lambda_{np}(t-t')}{2\epsilon} \right] R_{np}(\xi)R_{np}(\xi') 
\\
\nonumber
&&
\times \Big[ Y^c_{nm}(\theta, \phi) Y^c_{nm}(\theta', \phi')+ Y^s_{nm}(\theta, \phi) Y^s_{nm}(\theta', \phi')\Big]
\ .
\eeqn
The $Y_{nm}$ are the normalized spherical harmonics defined as:
\beqn
\left.
\begin{array}{l}
Y^c_{nm}(\theta, \phi)
\\
Y^s_{nm}(\theta, \phi)
\end{array}
\right\}
&=&
\sqrt{\frac{(2n+1) (n-m)!}{2\pi (1+\delta_{0m}) (n+m)!}} 
 \ P_n^m(\cos \theta)\left\{
\begin{array}{l}
\cos m \phi
\\
\sin m \phi \ ,
\end{array}
\right.
\eeqn
where $P_n^m$ are the Legendre polynomials. Furthermore, in \eq (\ref{Gf}), $R_{np}$ are defined by the following eigenvalue equation with eigenvalue $\lambda_{np}$:
\beq
\frac{1}{\xi} \frac{\pp^2}{\pp \xi^2} (\xi R_{np}) - \frac{n(n+1)}{\xi^2} R_{np} 
=\left[\epsilon \kappa (\xi-1)^2-\lambda_{np} \right] R_{np}
\ .
\eeq

Now, the $s$-mer partition function can be computed by integrating over all the possible choices for the two end points in the above Green's function. As a result,
\beqn
\nonumber
Z_s&=&\ee^{-(s-\gamma)\mu}V\int \dd \theta \dd \phi \dd \xi \dd \theta' \dd \phi' \dd \xi'
\xi^2 \xi'^2\sin \theta \sin \theta'  G(L_s,\xi, \theta, \phi| 0,\xi', \theta', \phi')
\\
\label{Zres}
&=& 4\pi\ee^{-(s-\gamma)\mu}V\sum_{p=0}^\infty  C_p\exp\left[-\sqrt{\frac{\kappa}{\epsilon}}\frac{ (2p+1)s}{2} \right]
\eeqn
where the expression in \eq (\ref{Zres}) is valid in the limit $\epsilon\kappa \gg 1$ \cite{saito}, and $C_p$ are defined as \cite{saito,Gradshteyn_B07}
\beq
C_p =
\left\{
\begin{array}{ll}
\frac{\sqrt{\pi}}{(\kappa \epsilon)^{1/4}} \frac{2p!}{2^p[(p/2)!]^2} \ ,& p \ \ {\rm even}
\\
2\sqrt{\frac{2\pi}{\kappa \epsilon}} \frac{p!}{2^p[(p-1)/2]!} \ ,& p \ \ {\rm odd} \ .
\end{array}
\right.
\eeq
Employing \eq (\ref{phis}), the length distribution is then
\beq
\label{new_res}
\phi_s \propto \ee^{- s/\alpha }\left[ 1+ \left(\frac{\kappa \epsilon }{\pi^2} \right)^{1/4}
\sum_{p>1}C_p \ee^{-\sqrt{\frac{\kappa }{\epsilon}} p s} \right]
 \ .
\eeq
Due to the series summation in the square brackets, the constant $\alpha$, set by the total concentration of the monomers in the system, no longer corresponds to the average length of the polymers. Furthermore, since $C_p >0$ for all $p$, the effects of the bending and stretching of the polymers are to make the length distribution non-exponential (\cf \fig \ref{plot}). In particular, in comparison to the mean-field prediction, the standard deviation in the length distribution of the polymers is decreased.

Indeed, the summation in the square brackets constitutes the correction to the mean-field prediction. Physically, the correction terms stem from the ability of a  short polymer to sense the boundary effects  due to the rigidity (the term proportional to $\epsilon$ in \eq \ref{stiffE}). Intuitively, this effect will disappear as the length increases, which is the case 
since the terms in the series are rapidly decreasing due to the form of $C_p$ and the exponential terms. In other words, the length distribution should be well described by the exponential distribution when the average length is much greater than 1.

\section{Relevance to amyloid fibrils}
\label{discussion}
In the case of self-assembled amyloid fibrils,
experimental measurements on the length distribution has been carried by various groups %%@
\cite{Rogers_Macromol05,Rogers_EPJE05,vanRaaij_BiophysJ08,Xue_JBC09,Xue_ProtEngDesignSelect09}. 
 A typical amyloid fibril has a persistence length in the order of $1-10$ $\mu$m  and has an average length in the order of 1 $\mu$m in typical 
experimental conditions \cite{knowles2,adamcik,castro,knowles1}. Various amyloid fibrils have also been shown to share similar mechanical properties  
\cite{knowles1}. In particular, the Young's modulus and bending rigidity for the
insulin amyloid fibrils were found to be in the order of 3 GPa and $10^{-25}$ Nm$^2$ respectively \cite{knowles2}. In our non-dimensionalised units, where $k_BT$ is set to one and the unit of length corresponds to the average fibril length per insulin protein, which is in the order 1 nm \cite{knowles2}, one finds that $\kappa \simeq \epsilon \simeq 10^{4} $.  Therefore, we expect that our analysis performed here is appropriate for the study of amyloid fibrillisation. Specifically, we expect that the fibrillar length distribution at thermal equilibrium is well described by the exponential form. Experimentally, all measured length distributions seem to show broad distribution and exponential-like %%@
decay in the tail. This correlates well with the prediction. 
On the other hand, contrary to our prediction,
peaks in the length distributions have also been observed in some cases \cite{ 
vanRaaij_BiophysJ08,Xue_JBC09}. van Raaij {\it et al.} has interpreted the observed peaks as a result of the finite resolution of the 
atomic force microscopy imaging and length measurement procedure \cite{vanRaaij_BiophysJ08}. Besides this explanation, it is also known that it can take in the order of %%@
months for mature fibrils to form \cite{Morel_BiophysJ10}. Therefore, the appearance of the peaks observed may also reflect the fact that the self-assembled systems have not yet reached thermal equilibrium 
due to the kinetic
barrier in the  nucleation process \cite{lee2,Lomakin_PNAS97, Auer_PRL08,Cabriolu_JChemPhys10}.

\section{Conclusion}
In summary, we have studied 
length distribution of self-assembled,  long and stiff polymers.  Our analysis is based on the calculation of the total free energy %%@
of the system and our conclusion is that at thermal equilibrium, the length distribution is well described by the exponential distribution, except in the short length limit. 
We have also discussed the implications of our results on the experimentally observed length distributions in
amyloid fibrils.

The two main limitations of our investigations are the followings. First, our analysis applies only to self-assembled stiff polymers at thermal equilibrium. Indeed, for nonequilibrium systems, non-exponential length distribution can occur \cite{kruse, frey}. Second, our approach applies strictly to the dilute limit where pairwise polymer interactions are negligible. 
As the concentration of polymers increases, the pairwise interactions will become important. If the interactions are purely steric, then an isotropic-nematic transition will occur as concentration increases. In this case, according to mean-field theory,  the length distribution of the system will remain exponential although the average length will increase drastically as the transition occurs \cite{odijk,schoot2,lee1}. Since our analysis suggests that mean-field theory is a good approximation for self-assembled stiff polymers in the dilute limit, it may be reasonable to expect that the mean-field prediction will continue to hold in this scenario as well.

\ack
%\acknowledgments
The author performed part of this work at the Max Planck Institute for the Physics Complex Systems in Dresden. He also thanks Frank J\"{u}licher  for helpful comments.

\end{document}